\documentclass[epj]{svjour}
\usepackage{graphicx}
\begin{document}
\title{Collective oscillations in quantum rings: 
a broken symmetry case}
\author{M. Val\'{\i}n-Rodr\'{\i}guez \and 
        A. Puente \and 
        Ll.\ Serra}
\institute{Departament de F\'{\i}sica,
Universitat de les Illes Balears, E-07071 Palma de Mallorca, Spain}
\date{24 march 2000}

\abstract{
We present calculations within density functional theory of the 
ground state and collective electronic oscillations in small 
two-dimensional quantum rings. 
No spatial symmetries are imposed to the solutions
and, as in a recent contribution, a transition to a broken
symmetry solution in the intrinsic reference frame for an increasingly 
narrow ring is found. The oscillations
are addressed by using real-time simulation. 
Conspicuous effects of the broken symmetry solution on the 
spectra are pointed out.
} 
\maketitle

\section{Introduction}

Quantum rings have attracted much interest in recent years. They 
constitute a clear example of the great technological advances in 
the fabrication of electronic nanostructures, which also include 
a rich variety of quantum dots. Theoretical studies of rings have addressed 
the appearance of the so called `persistent current' as a function of
an applied magnetic field, as well as the electronic structure and 
optical excitations within a fully microscopic approach for small 
rings \cite{Chak94,Chak96,Wend96}.
On the other hand, 
semiclassical methods have been applied to larger rings \cite{Zar96},
for which there 
exist experimental results in the mesoscopic domain \cite{Dha93}. 
These semiclassical methods have indeed provided a good
description of the $B$-dispersion in the measured far-infrared collective 
excitations for large rings. At an intermediate theoretical level, microcopic 
density-functional calculations have been applied to describe rings 
containing around $N\approx10$ electrons \cite{Rei99,Emp99}. 
An exciting prediction 
of these latter calculations is the transition for increasingly narrow 
rings (approaching the quasi-1d limit) to a state with broken rotational 
symmetry in the intrinsic reference frame
containing a static spin-density wave (SDW).
The fabrication of such rings, with a small number of electrons, 
is a very demanding challenge that only recently has been 
addressed by Lorke {\em et al.}, who reported 
measurements of the far infrared response of two electron 
rings \cite{Lor99,Lor98}.

All calculations of ring excitations for intermediate sizes ($N\geq 10$),
using either semiclassical methods or density-functional theory (DFT), 
have imposed circular symmetry to the system and, therefore, have not
been able to signal the influence of the broken symmetry ground state
on the spectra. In this paper we relax the symmetry condition on the 
ground and excited states of rings with the purpose to explore 
these effects within DFT. 
To this end we will use real time simulations of the collective density 
oscillations developed by two of us in the context of quantum 
dots \cite{Pue99,Ser99}.
The paper is organized as follows:
in Sec.\ II we consider the ring ground-state within our approach,
focussing on the transition to the symmetry-broken ground state;
in Sec.\ III we address the real time simulation of the charge and
spin-density oscillations. Finally, the conclusions are drawn in 
Sec.\ IV.

\section{DFT description of rings}

In order to model two-dimensional (2d) quantum rings of varying widths 
we have used the model proposed by Reimann {\em et al.} \cite{Rei99} 
and consider a ring confining potential on the $xy$ plane depending only 
on the radial distance $r$, such as
\begin{equation}
V^{({\mathit ext})}(r)={1\over 2} {\omega_0^2\over\delta} (r-R_0)^2\; .
\end{equation}
In this expression $R_0$ corresponds to the ring radius and we take 
$\omega_0^2=1/r_s^3N_p^{1/2}$, with $N_p$ and $r_s$ parameters analogous
to those giving for a circular quantum dot the number of positive 
charges and the Wigner-Seitz radius, respectively. The $\delta$ parameter 
is introduced in order to model the radial thickness of the 
ring. As delta decreases, the ring becomes thinner and thinner,
approaching the quasi-1d limit. 

We describe the electronic structure within 
the local-spin-density approximation of DFT. The reader is referred to 
Refs.\ \cite{Kos97,Pi98}
for details of the approach in the present context. 
The set of single-particle (sp) orbitals is obtained by selfconsistently 
solving the Kohn-Sham (KS) equations
\begin{equation}
\label{eq2}
\left[
-{1\over 2}\nabla^2 + V^{({\mathit ext})}(r)+ 
V^{(H)}({\bf r}) + V_\eta^{(xc)}({\bf r})
\right]
\varphi_{i\eta}({\bf r}) = \varepsilon_{i\eta} \varphi_{i\eta}({\bf r})\; ,
\end{equation}
where $V^{(H)}$ and $V_\eta^{(xc)}$ are respectively the Hartree and 
exchange-correlation potentials. The latter one is obtained from the 
local energy density  as
\begin{equation}
V_\eta^{(xc)} ={\partial {\cal E}_{xc}(\rho,m)\over\partial\rho_\eta}\; ,
\end{equation}
In these expressions $\eta=\uparrow,\downarrow$ labels the spin components
while the spin densities are 
$\rho_\eta({\bf r})=\sum_i{|\varphi_{i\eta}({\bf r})|^2}$. Total density 
and spin magnetization are given by 
$\rho=\rho_\uparrow+\rho_\downarrow$ and
$m=\rho_\uparrow-\rho_\downarrow$, respectively. Our exchange-correlation 
functional is based on the Tanatar-Ceperley results for the 2d electron 
gas \cite{Tan89}
with the von Barth-Hedin interpolation for intermediate 
polarizations \cite{Bar72}.

We emphasize here that our solution to Eq.\ (\ref{eq2}) is not assuming
{\em a priori} any symmetry for the orbitals $\varphi({\bf r})$. 
The 2d plane is discretized into a uniform grid of points in Cartesian 
coordinates and the Laplacian operator is approximated by the corresponding 
finite differences (we have used typically 7 point formulas).
The KS solutions are then obtained by iteratively applying the 
{\em imaginary-time-step} method.
This technique is very robust although convergence may be rather 
slow in some cases. The lowest energy solution
in each case is obtained by performing calculations with different 
starting points in order to assure that the result does not correspond 
to a local minimum representing an excited state of the system.

\subsection{Circular rings}

Figure 1 shows the density and magnetization for a ring with \cite{units} 
$R_0=8$ , 
containing $N=10$ electrons and for two values of the 
width $\delta$. The wide one ($\delta=2$) has perfect circular symmetry
and its magnetization vanishes. On the contrary, the narrow one 
($\delta=0.08$) presents a modulation in charge density, a charge 
density wave (CDW), and spin polarization of antiferromagnetic type,
with alternate orientations of spin up and down (a SDW).
This drammatic change of behaviour is associated with the increase of the
radial confinement and agrees with the findings of Ref.\ \cite{Rei99}.
We compare in Fig.\ 2 the energy of the unrestricted solution 
to Eq.\ (\ref{eq2})
with that obtained with the contraint of circular symmetry, i.e.,  
by solving only the radial KS equation as in Ref.\ \cite{Emp99}.
Also shown in Fig.\ 2 is the ratio of maximum magnetization to maximum 
density and the amplitude of the CDW.
Panels (b) and (c) show that the transition is rather abrupt with the value
of $\delta$, although the energy (panel (a)) is smoothly changing.
Nevertheless, the energy trend is very clear and manifests the gain 
in binding energy with the formation of CDW's and SDW's.

Although some controversy has existed in the quantum dot
community \cite{Kos97,Hi99,Yan99},
the formation of broken-symmetry solutions as those displayed 
by mean-field theories is a well known phenomenon
in nuclear physics \cite{Rin90} and in atomic physics \cite{Ber89}. 
It has been recently discused in the context 
of 2d quantum dots by Yannouleas and Landman \cite{Yan99}
and by Koskinen et al \cite{Kos97,Kos00}.
It has been shown that 
the mean field solution corresponds to the intrinsic structure of the 
system although the full exact solution in the laboratory frame preserves 
the symmetry because of an underlying degeneracy. 
This is convincingly shown by comparing internal 
structure properties of the exact solution, such as the conditional pair 
probability, with the mean field result\cite{Yan99}. 
Another strong evidence of the internal symmetry breaking is given 
by the comparison of the low lying states with the roto-vibrational 
states of the structure formed by the localized electrons
\cite{Yan99,Kos00}.
Quantum rings constitute indeed a very good 
scenario for this behaviour since, as shown in Figs.\ 1 and 2, they 
exhibit strong breaking of the circular symmetry. A critical comparison of 
the mean field solution with the exact one for two-electron quantum rings
will be presented in a future publication.

The broken-symmetry solution is accompanied by a 
strong change in the sp level scheme. While 
the circular structure is characterized by shells, with closings 
at electron numbers of 2, 6, 10, 14 and a small sensitivity 
on the width $\delta$,
the formation of SDW's and CDW's leads to 
a bunching of the occupied levels and the formation of a relatively
large energy gap at the Fermi level. This behaviour is shown in 
panel (d) of Fig.\ 2.

\subsection{Elliptic rings}

Deformed rings can be simulated by using an elliptic contour for $R_0(\theta)$
of the type
\begin{equation}
R_0(\theta)= a \left(\cos^2\theta+ {1 \over \beta^2} \sin^2\theta 
\right)^{-{1\over 2}} \; ,
\end{equation}
where $a$ is the major axis of the ellipse, $\beta$ gives the ratio 
of minor to major axis and $\theta$ is the polar angle.
In order to compare with the circular rings 
of the preceding subsection we have taken elliptic rings 
with the same perimeter and have considered two deformations, namely
$\beta=0.75$ and $\beta=0.5$, corresponding to medium and large 
ring deformation, respectively.

One of the most striking findings in elliptic rings is the appearance 
of the broken-symmetry solution along the ring contour {\em for 
increasing deformation}, at fixed width.  
This happens even for wide rings  
although the wave structure is more clearly marked in the narrow ones
(Fig.\ 3). Another behaviour which is apparent in Fig.\ 3 is the 
formation of charge concentrations at the ends of the long axis.
These are attributed to {\em end states} similar to those found 
in finite wires \cite{Rei99}. In broad rings, the formation of end 
states appears
abruptly after a minimum deformation is reached. For instance, 
at $\delta=2$ and $\beta=0.75$ no evidence is yet found of their 
formation.

\section{Time dependent spin and charge density oscillations}

The analysis of excited states is often made by invoking
perturbation theory for an external probing field. The feasibility 
of this approach usually relies on the symmetries of the system, 
which greatly simplify the treatment. For instance, in circular dots one
can take density oscillations of the type 
$\delta\rho_\eta(r) e^{im\theta}$, i.e., a radial function
times a multipolar field, and then reduce the equations to the
radial parts. In the present context the possibility
of having a broken symmetry ground state forbids this approach. 

A suitable alternative to perturbation theory is to study the density 
oscillations by using real-time simulations. This is particularly well
adapted to DFT and leads to its time-dependent 
generalization (TDDFT). The simplest version of this scheme is the 
adiabatic-local-density approximation, which uses the energy functional 
of the ground state 
to explore small amplitude oscillations around it. This is one of the 
simplest versions of TDDFT and yet it is quite robust and satisfies 
known exact properties as the energy weighted sum rule and the 
generalized harmonic-potential theorem \cite{Dob94,Lipp97}.
We will use it to calculate 
the ring spin-density oscillations.

The time-dependent Schr\"odinger equations can be written
\begin{equation}
\label{eqt}
i{\partial\over\partial t}\varphi_{i\eta}({\bf r},t) =
h_\eta\left[\rho,m\right] \varphi_{i\eta}({\bf r},t)\; ,
\end{equation}
where the KS single-particle Hamiltonian $h_\eta\left[\rho,m\right]$ is given
by the square bracket in Eq.\ (\ref{eq2}), whose 
stationary solution was found in the preceding Section. Here we
will perform an initial (at $t=0$) pertubation of the wave functions 
\begin{equation}
\varphi_{i\eta}({\bf r},0)={\cal P} \varphi_{i\eta}({\bf r})\; ,
\end{equation}  
and track the time evolution through Eq.\ (\ref{eqt}). 
Technical details of the 
integration method, as well as of the analysis of the dipole 
signals $\langle {\bf r}\rangle_\eta$ which allow to extract
the excitation spectrum can be found in Refs.\ \cite{Pue99,Ser99}.

We will distinguish three different types of excitation modes, 
two of them depending on the initial perturbation ${\cal P}$ and 
another one corresponding to a theoretical model in which the electrons 
are not interacting. Namely: {\em density modes} associated with an 
initial rigid displacement of the total density; {\em spin modes}
for which at $t=0$ the spin densities are rigidly shifted in opposite
directions; and {\em free modes} for which the sp Hamiltonian
is kept fixed to the stationary densitites, i.e., 
$h_\eta\left[\rho,m\right]\approx h_\eta\left[\rho_0,m_0\right]$. 
The free oscillations, also known as sp modes,
are equally excited by shifting either the total density 
or the spin densities.

\subsection{circular rings}

Figure 4 shows the free, density and spin spectra for the broad ring with 
$N=10$ and $\delta=2$. As known from other 
calculations \cite{Zar96,Emp99} the spectrum
is roughly divided in two regions in the sp model which are then 
shifted by the interaction in the density and spin channels. The lower 
region contains a single peak while the higher one exhibits 
an important fragmentation.
The density response is characterized by the blue shift from the sp peaks 
while the spin response exhibits a shift to lower energy. The residual 
interaction is also quite effective in enhancing one peak above the others
in the high energy region. The peak that collects the strength
is then associated to the more collective excitation. 
These features can be 
considered standard in finite Fermi systems with a repulsive interaction
in the density channel and an attractive one for the spin channel. Indeed, 
a similar behaviour has been found in quantum dots. 
Figure 4 also shows the 
corresponding spectra obtained by using the circularly constrained solution
in the perturbative random-phase approximation (RPA) \cite{Emp99}. It is 
obvious that both results are equivalent, as one would expect since the 
ground state has circular symmetry. We attribute the minor 
diferences to the quite different techniques that have been used in the 
two cases. It is worth to notice that the real time method is much more
demanding computationally and that very fine details of the spectrum 
are harder to describe, such as very low intensity peaks, or the resolution
of two closely lying excitations.

The results for the narrow ring ($\delta=0.08$) are shown in Fig\ 5. 
Qualitative differences appear with respect to the preceding ring and,
most important, with respect to the circularly constrained solution. 
They must be attributed to the symmetry breaking ground state of this ring.
The separation of high and low energy region is now clearer, with an important 
gap between both. The most striking result comes from the 
comparison of the low energy peaks with the circularly contrained solution.
First we notice that the full solution in this energy region yields a 
density response which is very close to, but just below, the sp peaks 
and overlaps with some excitations of the spin response.
In the circularly contrained calculation the density response is above the
sp one and, furthermore, the spin peak lies at an essentially vanishing energy.
The different behaviour is indicating that the symmetry breaking is 
giving stability to the ring, which would otherwise become 
unstable against spin oscillations. This is in agreement with the appearance
of an energy gap in the sp energies associated to the symmetry breaking,
mentioned in Sec.\ II.

The high energy region is associated to a locally radial oscillation in the 
ring, and thus similar to the oscillation in a one dimensional
oscillator. This is confirmed by the similarity of the energy with the 
radial curvature at the minimum given by $\omega_0/\sqrt{\delta}$ that, 
by the well known generalized Kohn's theorem, is the only allowed energy for 
rigid density oscillations of dipole type in parabolic confinement
\cite{Mak90}. 
On the other hand, the low energy 
region is associated with oscillations locally tangential to the ring.
Within this interpretation, the fact that the density 
and spin tangential modes are close in energy
and essentially overlap with the
sp excitations can be understood as a signature of a 1-dimensional
Luttinger liquid \cite{Hal81}, 
for which the lowest excitations are not the sp but the collective ones.

In the symmetry broken ground state there is also a possibility to 
excite pure tangential modes by means of a twist of the two spin densities
in opposite directions. These {\em exotic} modes were proposed recently 
in the context of quantum dots \cite{Pue99}. 
It is obvious that pure tangential 
modes can not exist in circularly symmetric rings since a twist of 
spin densities 
does not change the energy. However, in a symmetry broken state
a restoring force appears because of the distortion of the spin-density 
wave. The associated oscillation energy is easily obtained by 
analyzing the time dependence of the orbital currents 
$\langle \sum_i{\ell_i^{(z)}\sigma_i^{(z)}}\rangle$ that appear after 
an initial spin twist. 

Figure 6 shows the spin twist spectrum for the narrow ring with 
10 electrons. We observe that the peaks are in the same 
region as the low energy dipole excitations, which is proving 
that the latter ones are indeed of tangential nature.
Both free and interacting spectra are characterized by a 
rather regular energy spacing, after an initial gap.
We attribute this to the parabolicity of the potential in the 
tangential direction for points close to the local minima.

\subsection{Elliptic rings}

The systematics with deformation is very similar to that discussed for
the circular case, with the additional result that both dipole energy regions 
are now more fragmented. 
Figure 7 shows the corresponding spectra 
for a strongly deformed ring $\beta=0.5$ with thicknesses given by
$\delta=2$ and $\delta=0.08$. This fragmentation is more regular 
in the wide rings, with a doubling of high and low energy peaks 
by the same energy separation. In the narrow ring the relative fragmentation 
is lower with respect to the separation of high and low energy regions which 
is in agreement with the previously mentioned transverse and longitudinal 
character of both regions.

\section{Conclusions}

The ground state and oscillations of quantum rings have been 
discussed within density-functional theory relaxing the constraint 
of circular symmetry of the electronic density. A transition to a broken 
symmetry ground state with static charge and spin density waves has been 
found for increasingly narrow rings, in agreement with previous results.
The energy gain with respect to the circularly contraint solution, the 
amplitude of charge and spin-density wave as well as the sp 
level scheme have been discussed as a function of ring width.
The transition to a wave structure along the ring contour 
has also been obtained in 
elliptic rings for increasing deformations. In this case, there also 
appear charge concentrations at the long axis ends indicating the 
formation of end states similar to those obtained in finite quantum
wires.

Signatures of the broken symmetry ground state on the collective 
oscillation energies have been searched for in the dipole spin and density 
oscillations. We have shown that the broken symmetry is accompanied by 
an overlap of the density and spin response peaks in the low energy region
corresponding to tangential oscillations. At the same time, the 
low lying sp excitations are at a slightly higher energy, which is indicating
a quasi 1d behaviour similar to that of Luttinger liquids. The energy of
these excitations is close to the gap associated to the broken symmetry.
It has also been shown that the broken symmetry gives stability to 
the ring, which would otherwise become unstable against spin dipole 
oscillations. 
The oscillations in the radial direction are located at higher energy, close 
to the radial curvature of the parabola as indicated by the generalized 
Kohn's theorem. The separation between high and low energy regions
incresases as the ring width decreases. 

Pure tangential modes in rings with broken symmetry, associated to a twist 
of the spin densities have been discussed and it has been shown that their 
energy is close to the low dipole peaks. This supports the interpretation
of this dipole excitations. Besides, we have shown that the more exotic
spin-twist modes are regularly spaced in energy because of a high degree 
of harmonicity of the potential. Finally, the dipole oscillations in 
elliptic rings have been obtained and their fragmentation depending 
of the ring width has been shown.
 
Further work is necessary in order to analyze in detail the convergence 
to the strict 1d limit, for very narrow rings, and how the peculiarities 
of a Luttinger liquid such as power-law (as opposite to long range) order
in the SDW correlation function and spin-charge separation 
\cite{Hal81} may manifest in a finite size system.

\section{Acknowledgements}

M.V.-R.\ gratefully acknowledges support by the Consell de Mallorca-UIB.
This work was performed under grant No.\ PB98-0124.
%

%
\begin{figure}[h]
\caption{Results for a ring with $N=10$ electrons, $R_0=8$
and parameters $N_p=10$, $r_s=1.51$ (see text). 
From left to right the density for a ring width $\delta=2$,
the density for $\delta=0.08$ and the magnetization for
$\delta=0.08$ are shown, respectively.}
\label{fig1}
\end{figure}
\begin{figure}[f]
\includegraphics*[width=3.5in]{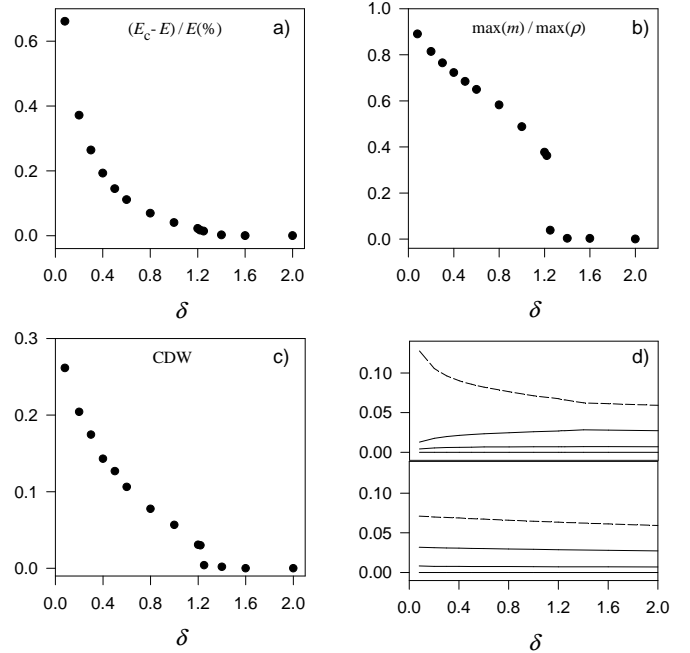}
\caption{Evolution with $\delta$ of several properties for the same ring of
Fig. 1. Panel (a) shows $(E_c-E)/E$ in percentage,
where $E_c$  is the energy of the circularly contrained
solution while $E$ is that of the full solution to Eq.\ (2). 
Panel (b) shows the evolution of the ratio of maximum magnetization over 
maximum density. Panel (c) displays the amplitude of the charge-density 
wave and panel (d) shows the evolution of the energy level scheme
referred to the lowest sp orbital. 
In this latter panel, occupied levels are plotted with solid lines while
the first unoccupied one is shown with a dashed line.}
\label{fig2}
\end{figure}
\begin{figure}[h]
\caption{Density and magnetization for elliptic rings with 10 electrons
and $\delta=2$. For $\beta=0.75$ only the density is shown,
since the magnetization vanishes, while for $\beta=0.5$
both density and magnetization are displayed.}
\label{fig3}
\end{figure}
%
\begin{figure}[f]
\includegraphics*[width=3.5in]{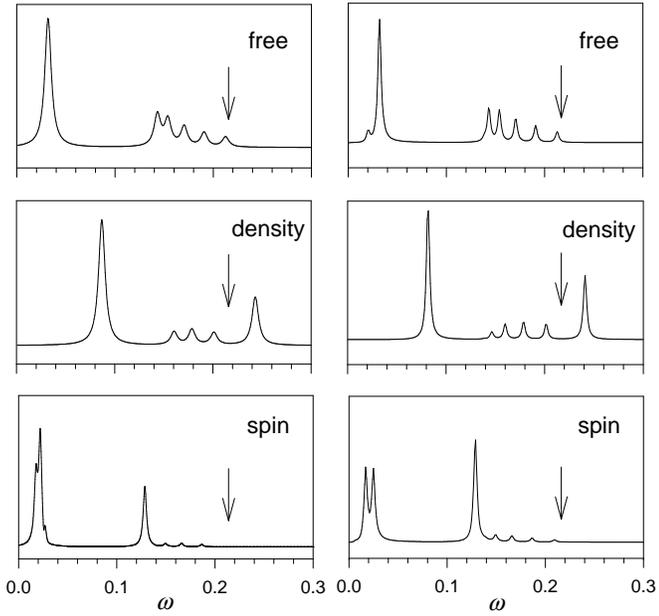}
\caption{Dipole oscillation spectra in the free, density 
and spin channels for the circular ring with $N=10$ and $\delta=2$.
The left panels correspond to the unrestricted solution while the right 
panels show the circularly constrained one.
The arrow shows the value $\omega_0/\sqrt{\delta}$, that 
gives the curvature of the radial parabola at the minimum.}
\label{fig4}
\end{figure}
\begin{figure}[f]
\includegraphics*[width=3.5in]{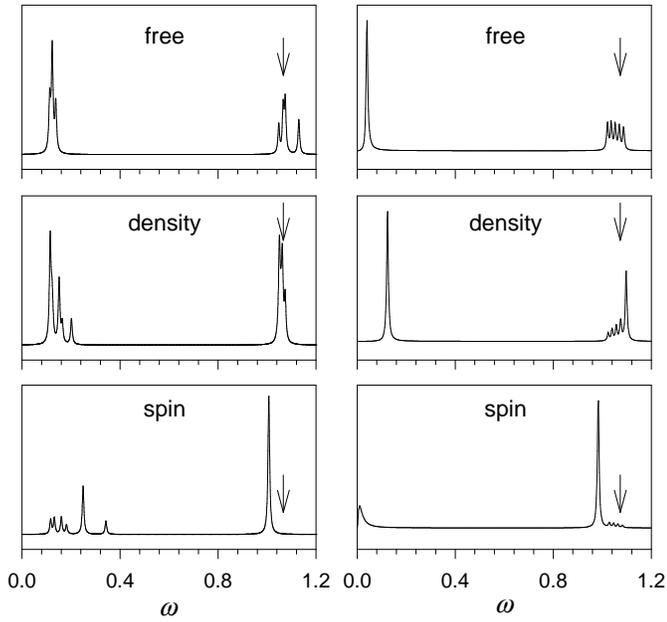}
\caption{Same as Fig.\ 4 but for $\delta=0.08$.}
\label{fig5}
\end{figure}
\begin{figure}[f]
\includegraphics*[width=2.8in]{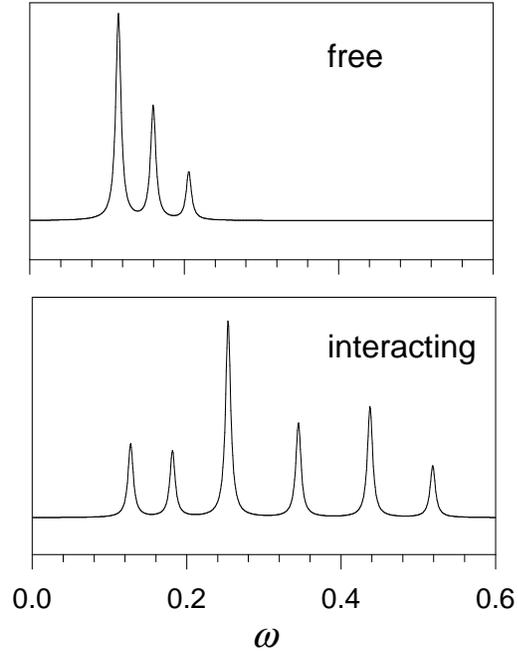}
\caption{Spin twist spectra for the narrow circular ring in the 
free and interacting channels.}
\label{fig6}
\end{figure}
\begin{figure}[f]
\includegraphics*[width=3.5in]{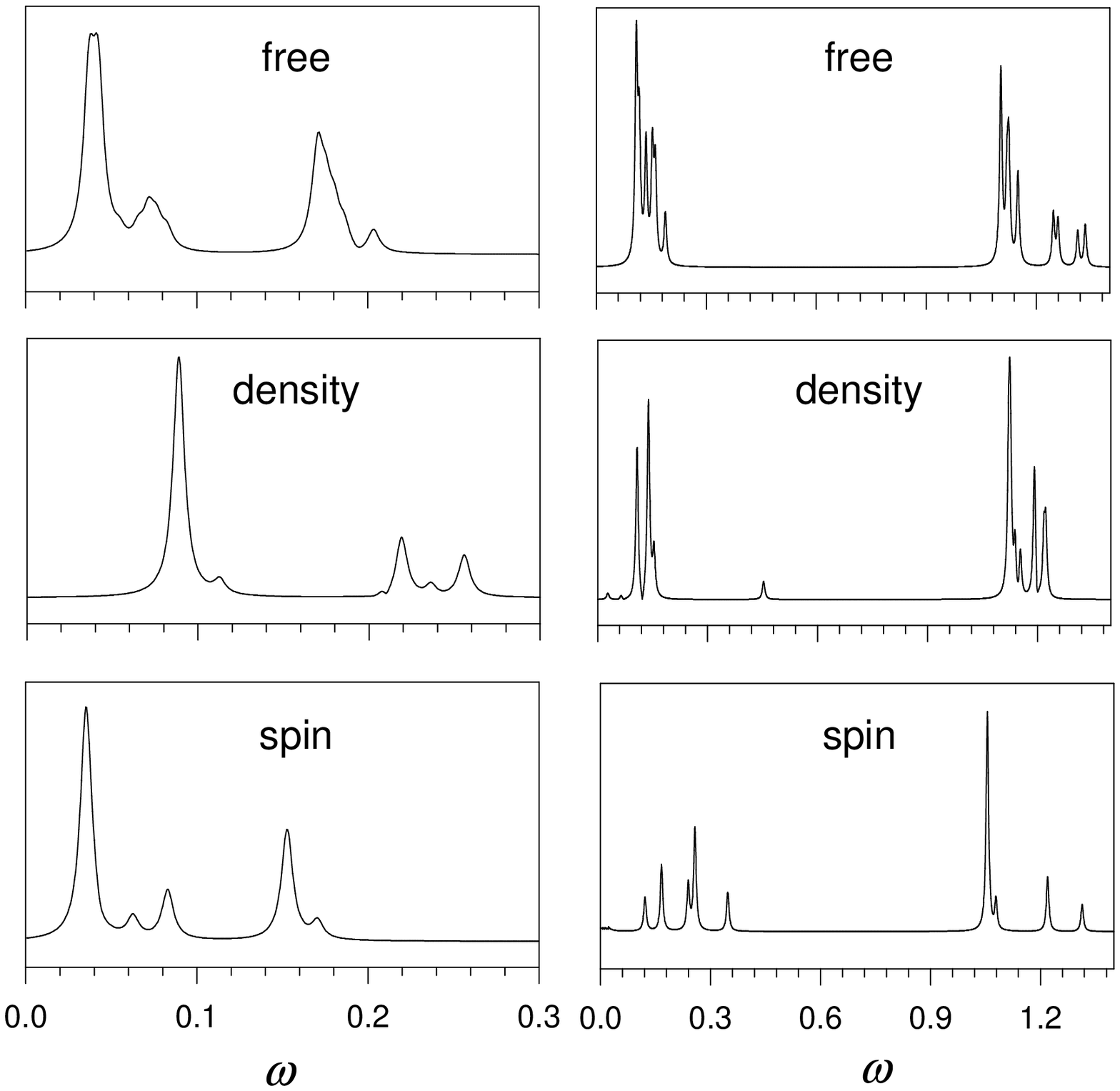}
\caption{Same as Fig.\ 4 for a deformed ring with $\beta=0.5$. Left 
panels correspond to $\delta=2$ and right ones to $\delta=0.08$.}
\label{fig7}
\end{figure}

\begin{thebibliography}{}
\bibitem{units} All numerical values, unless stated otherwise, are 
given in effective units, for which $\hbar=m^*=e^2/\epsilon=1$. 
Then, the energy unit is ${\rm H}^*=12$~meV and the lenght unit 
$a_0^*=98$~\AA.
\end{thebibliography}
\end{document}